\documentclass{article}
\usepackage{graphicx}
\usepackage{subcaption}
\usepackage{subcaption}
\usepackage{booktabs}
\usepackage{spconf,amsmath,graphicx}
\usepackage{float}
\usepackage{tabularx}
\usepackage[colorinlistoftodos,prependcaption,textsize=small]{todonotes} \presetkeys{todonotes}{inline,backgroundcolor=cyan}{}

\title{Training speech emotion classifier without categorical annotations}

\name{Meysam Shamsi, Marie Tahon}
\address{LIUM\\
	Le Mans University\\
	Avenue Olivier Messiaen,72085 Le Mans, France}
\begin{document}
%\ninept
%
\maketitle
\begin{abstract}
% intro
There are two paradigms of emotion representation, categorical labeling and dimensional description in continuous space. 
Therefore, the emotion recognition task can be treated as a classification or regression.
% problem
The main aim of this study is to investigate the relation between these two representations and propose a classification pipeline that uses only dimensional annotation.
% idea
The proposed approach contains a regressor model which is trained to predict a vector of continuous values in dimensional representation for given speech audio.
The output of this model can be interpreted as an emotional category using a mapping algorithm.
% experiment
We investigated the performances of a combination of three feature extractors, three neural network architectures, and three mapping algorithms on two different corpora.
Our study shows the advantages and limitations of the classification via regression approach.
\end{abstract}
\begin{keywords}
Speech emotion recognition, Emotion representation, Classification, Regression
\end{keywords}
\section{INTRODUCTION}

The importance of extracting the paralinguistic information from speech has led the research community into Speech Emotion Recognition (SER).
But the definition of emotions is ambiguous \cite{cowie2003describing}.
% It makes a disagreement
Consequently, there is no consensus on emotion representation and annotation.
The two main emotional theories used in computer science are the followings: emotions can be described with categorical labels mostly based on Ekman representations \cite{ekman1992argument} or emotional dimensions such 
% Two main representations of emotion are categorical and attributes based.
% While categorical representation uses labels such as
% % neutral, 
% happy, sad, angry etc. for naming an emotional state, the attribute based representation use continuous values in different dimensions such
as arousal (or activation), valence, dominance (AVD) \cite{russell1977evidence} to precise the emotional state.

These two representations have merits and disadvantages.
Usually, using categorical labels for describing emotional states would be more understandable for the public \cite{gunes2011emotion}.
But it makes the representation of emotional states limited to certain categories, which may not cover all human emotions.
On the other hand, using continuous value can precisely assign the emotional state to a point in dimensional space, which is less close to human language.

From a machine learning point of view, the advantages of dimensional representation are in favor compared with categorical representation \cite{yannakakis2018ordinal}.
In the following, the benefits of using continuous values for emotions in dimensional space are detailed.
A supervised machine learning model typically uses ground truth annotation.
But due to the complexity of human emotions, there is always a disagreement on the perceived emotions and then annotations. 
So usually the assigned value of annotators would be aggregated to generate one single annotation per input.
One of the main differences between categorical representation, which makes emotion recognition a classification task, and the dimensional representation, which makes emotion recognition a regression task, is the conserved information after aggregation of annotations. 
The most commonly used method for the aggregation is getting the majority vote of the annotator's opinion to have a hard label.
Although, some studies such as \cite{lotfian2018predicting,ando2018soft,han2017hard} followed a soft labeling approach to deal with the labeling complexity and ambiguity.  
% However, some studies such as \cite{lotfian2018predicting,ando2018soft,han2017hard} profits from several categorical labels in the classification task, the most commonly used method for the aggregation is getting majority vote of annotators opinion to have a hard label.
%a soft labeling reference
For example, in the standard protocols of \textit{IEMOCAP} dataset \cite{busso2008iemocap} and \textit{MSP-Podcast} corpus \cite{lotfian2019msp}, the samples with disagreement of annotators are discarded.
% Besides wasting annotated samples with disagreement labels, using majority voting as an aggregation method would suppress certain labels of a sample on the categorical annotation.
% While getting average on the continuous value of dimensions keeps more information, however, their variance would be omitted.
% As one of the advantage of the dimensional annotation to the categorical annotation is that the majority voting suppress certain annotations on categorical emotions, but getting average on continuous value of dimensions keep more information, however their variance would be omitted.

The most common approach for encoding emotional categories is one hot vector, which ignores the relation or distance between emotions. 
For example, anger can be very close to irritation, frustration, and rage, and they are usually perceived or expressed in similar situations. 
On the contrary, dimensional annotation which provides a Distributed Representation can keep the intra and inter categories distance information. 
This continuous representation helps to break out the limitation of discrete labels.

It has been claimed by \cite{lotfian2019msp} that the perceived and expressed frequency of categorical emotions are not the same in the real life.
In this case, an imbalanced classification problem would be faced \cite{fujioka2020meta}.
% If the categorical representation have been used, an imbalanced classification problem would be faced in real-life applications \cite{fujioka2020meta}.
As the dispersion of emotional labels in different corpus has been shown in \cite{lotfian2019msp}, the impact of the non-homogeneous frequency of class would be less important when the dimensional representation is employed.

Last but not least, the dimensional representation has its application where a continuous value is needed more than only a category.
In the task of continuous emotion recognition\cite{gunes2013categorical,wollmer2008abandoning}, when a sequence of predictions over time is the goal, the dimensional approach helps to smooth the transition and take into account temporal dependencies.
% For example, the study of costumer satisfaction level can be interpreted from valance dimension with a higher level of precision.
All these conveniences emphasize on the capacity of dimensional representation of emotional states.

In this paper, the coherence of these two annotation types in two common used corpora, \textit{IEMOCAP} \cite{busso2008iemocap} and \textit{MSP-Podcast} \cite{lotfian2019msp}, is studied.
Moreover, the capacity of classification models without using categorical annotations is investigated. 
This approach can show the advantages of dimensional annotation and representation, which is theoretically and empirically supported in \cite{yannakakis2018ordinal}.

% \subsection{PREVIOUS WORKS}

% Using all annotators instead of only one majority vote label \cite{lotfian2018predicting}:
% using eGeMAPS (88 per utterance by statistics) / test set(7181 segments), development set (2614), train set (12835) >> best unweighted average F1-score: 26.3 (chance performance is 12.5)

% \cite{kowtha2020detecting} Regression model for attribute based emotion recognition, used more data than MSP Podcast for initial training, TC-LSTM model is used, study ROC curve to detect categorical emotion from Natural 

% \cite{lotfian2017formulating}: "the emotional perception of a stimuli is a multidimensional Gaussian random variable with an unobserved distribution. Each dimension corresponds to an emotion characterized by a numerical scale."

\section{THE CHALLENGE OF EMOTION ANNOTATION}
\label{sec_challenge}
% \todo{Intro to IEMOCAP and MSP-Podcast}

One of the main challenges of the emotional dataset creation and annotating human emotions is the annotator agreement.
As an example, two common datasets (\textit{IEMOCAP} and \textit{MSP-Podcast}) are compared in the following. 

Using hard labeling of categorical annotation, 19.4\% of samples in \textit{MSP-Podcast} corpus and 25\% of samples in \textit{IEMOCAP} could not get an agreed annotation from evaluators.
The reliabilities of agreement in these two corpora are not high.
The \textit{Fleiss’ Kappa} of categorical annotations in the \textit{MSP-Podcast} is only 0.23 and in the \textit{IEMOCAP} is 0.48.
A perceptual test in \cite{chernykh2017emotion} showed that the human performance for emotion recognition of four main classes in \textit{IEMOCAP} is only 69\% overall accuracy.
% \todo{\cite{lotfian2018predicting} human performance of MSP Podcast}

The evaluators' agreement on dimensional annotation is not high as well, except for the Arousal on the \textit{IEMOCAP}.
Table~\ref{tab:kri_a} shows the inter-evaluator reliability (\textit{Krippendorff's alpha} coefficient).
% More information about the experimental setting of these datasets can be found in section \ref{sec_datasets}.

\begin{table}[th]
  \caption{The Krippendorff's alpha coefficient of Arousal, Valance, Dominance annotations }
  \label{tab:kri_a}
  \centering
  \begin{tabular}{ c|c c c}
  \toprule
  Corpus & Arousal & Valance & Dominance \\
  \midrule
  \textit{IEMOCAP} & 0.68 & 0.30 & 0.27\\
  \textit{MSP-Podcast} & 0.27 & 0.33 & 0.21\\
  \bottomrule
  \end{tabular}
\end{table}

\begin{figure*}[]
    \centering
    \begin{subfigure}{0.95\columnwidth}
        \includegraphics[width=\columnwidth]{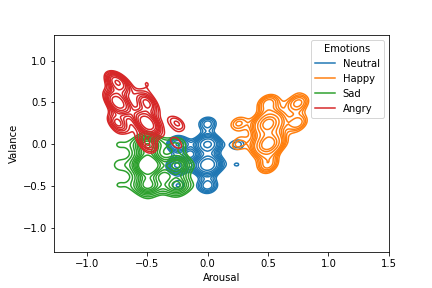}
        \caption{\textit{IEMOCAP}}
        \label{fig_imocap_map}
    \end{subfigure}
    \begin{subfigure}{0.95\columnwidth}
         \includegraphics[width=\columnwidth]{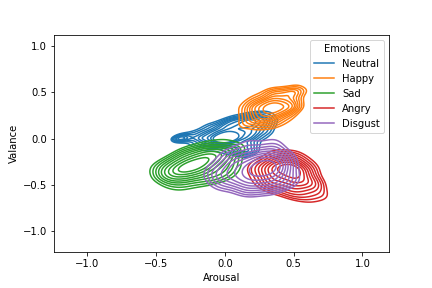}
        \caption{\textit{MSP-Podcast}}
        \label{fig_msp_map}
    \end{subfigure}
\caption{Density of the main emotion categories on normalized Arousal and Valence space. Non homogeneity of densities in \textit{IEMOCAP} is due to the limited number of annotators and scoring steps.}
\label{fig_Av2Class_map}
\end{figure*}

% 2. map emotions label on attribute based space > draw 2 figures of 2 datasets
To find a relation between two types of annotation, the density of samples with the same emotional label has been mapped in two main dimensions, arousal, and valence (See Figure \ref{fig_Av2Class_map}).
% As it has been suggested by \cite{lotfian2019curriculum,pappagari2020xvec} and 
In order to have comparable experimental results and as it has been suggested by \cite{lotfian2019curriculum,pappagari2020xvec}, only the 4 main emotions (\textit{Neutral}, \textit{Happy}, \textit{Sad} and \textit{Angry}) from \textit{IEMOCAP} and 5 main emotions (\textit{Neutral}, \textit{Happy}, \textit{Sad}, \textit{Angry} and \textit{Disgust}) from \textit{MSP-Podcast} have been used in the rest of this study.

Figure \ref{fig_Av2Class_map} confirms that the annotator in the different corpus has a different definition or perception of emotions. 
While the \textit{Angry} class in the \textit{IEMOCAP} has been evaluated with higher valence and lower arousal than \textit{Neutral}, it has been assigned to a lower arousal and higher valence values than \textit{Neutral} in the \textit{MSP-Podcast}.
This conflict indicates the ambiguity of emotions' definition, which makes the emotion classification challenging in the cross corpus cases.

\section{CLASSIFICATION VIA REGRESSOR}
\label{sec_class_via_reg}

\begin{figure}[H]
\centering
\includegraphics[width=\columnwidth]{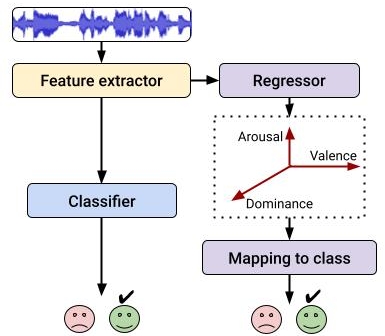}
\caption{Classification (the left pipeline) versus classification via regression (the right pipeline) for emotion recognition}

\label{fig_cla_via_reg}
\end{figure}

In order to investigate the relation between these categorical and dimensional annotations, the main goal of this study is to evaluate the ability of classification, based on the continuous values in the dimensional space.
Figure \ref{fig_cla_via_reg} demonstrates two pipelines for recognizing a class of emotion, a one-step classification on the left and classification via regression on the right.
The main idea is that the categorical label of samples can be predicted based on the dimensional values, as long as the annotations are consistence.
Some studies such as \cite{yannakakis2018ordinal,kowtha2020detecting} support the hypothesis.
In \cite{kowtha2020detecting}, it has been observed that a model for the prediction of arousal and valence values can be useful to detect categorical emotions.

Prediction of emotional categories based on a regression model on dimensional space has several benefits.
First, all labels in dimensional representation can be used in the training without wasting annotation costs.
It means not only samples with disagreement can be used in training, but also all annotations of a sample (by getting average of values instead of majority vote) can be taken into account.

Second, the problem of imbalanced data would be less important since the frequency of samples in different categories would not have a high impact on the training process in the dimensional representation.

Third, training on a distributed representation feature, which contains between and within class distances, can inject additional information into the training.

Forth, a trained regression model can be used for a classification task as well. 
In this case, based on the definition of categorical labels in dimensional space, the output of the regression model can be mapped to emotional vocabularies.
It means the parallel annotations, categorical and dimensional, of a dataset would not be necessary.
Only dimensional annotation and a mapping definition would be enough to have a prediction in two representations.
This potential is the main focus of this study's experiments.

For the mapping, three algorithms are proposed; Gaussian classifier (\textit{Gaussian}), K-Nearest Neighbors (\textit{KNN}) (optimized K=50) and Tow Layer Perceptrons, 5*5, (\textit{2LP}).
These models are constructed to predict the categorical label based on dimensional values.
In order to have an upper bound of classification performance based on dimensional values in three dimensions, the result of these mapping algorithms on reference annotation of \textit{IEMOCAP} and \textit{MSP-Podcast} is evaluated.
The results show that the valance is the most discriminative attribute (followed by arousal and dominance) to predict samples' labels in both corpora.
The combination of three attributes achieves the highest accuracy.
Table \ref{tab_dummy} compares three dummy models with proposed mapping algorithms which predicts the class label based on the ground truth values of \textit{arousal}, \textit{valance} and \textit{dominance}. 
The dummy models are \textit{Random} labels, \textit{Prob. Random} which generates labels randomly with respecting the probability of each class, \textit{Major Class} which always generates the label of the most frequent class.

It is important to compare the performance of a Machine Learning model with these dummy models, especially in an imbalanced dataset.
% In some applications, the frequency of classes as weight should be taken into account for any evaluation, especially in imbalanced data.
Based on the application, the evaluation metric can be different.
When the task is emotion recognition in a real-life situation, the weighted performance can be more important. 
On the other hand, when the task is only to distinguish between different emotions, the unweighted performance can be applied.
For this purpose, it is decided to report unweighted recall (UR) and weighted recall (WR).
Needless to mention, contrary to classification models, the imbalanced data can cause less impact on regression models' performance.
As it is noted in Table \ref{tab_dummy}, the weighted recall (WR) of selecting the only major class in the \textit{MSP-Podcast} is 50.6\%.

\begin{table}[th]
  \caption{Mapping performance from ground truth dimensions (arousal, valance and dominance) to categorical emotions }
  \label{tab_dummy}
  \centering
  \begin{tabular}{ c|c c|c c}
  \toprule
   & \multicolumn{2}{c}{\textit{IEMOCAP} (5fold)} & \multicolumn{2}{|c}{\textit{MSP-Podcast} test} \\
  Algorithm & UR & WR & UR & WR  \\
  \midrule
  Random & 25.0\% & 25.0\% & 20.0\% & 13.4\% \\
  Prob. Random  & 25.0\% & 26.2\% & 20.0\% & 36.0\% \\
  Major Class & 25.0\% & 33.5\% & 20.0\% & 50.6\% \\
  \midrule
  Gaussian & 71.2\% & 72.5\% & 54.6\% & 69.2\% \\
%   GNB(NoP) & 72.1\% & 72.5\% & 59.2\% & 56.1\% \\
  KNN & 68.8\% & 69.4\% & 53.0\% & 69.5\% \\
  2LP & 71.6\% & 72.7\% & 50.9\% & 72.3\% \\
  \bottomrule
  \end{tabular}
\end{table}

% IEMOCAP
% Random WA:25.01 UA:25.01
% Class probablity based WA:26.26 UA:24.99
% Major Class WA:33.5 UA:25.0
% MLP WA:72.71 UA:71.6
% KNN WA:69.38 UA:68.85
% GNB_noP WA:72.54 UA:72.07
% GNB_P WA:72.54 UA:71.18

%MSP-Podcast
% Random WA:12.37 UA:20.0
% Class probablity based WA:36.03 UA:20.0
% Major Class WA:50.51 UA:20.0
% MLP WA:71.32 UA:50.85
% KNN WA:69.51 UA:53.01
% GNB_noP WA:56.09 UA:59.22
% GNB_P WA:69.19 UA:54.62

The \textit{2LP} archives the best performance for the prediction of the emotional labels based on three attributes in the \textit{IEMOCAP} (with 4 classes). 
Although, it shows that a perfect regressor can map only 72.7\% of samples from AVD space to the classes of emotion.
The mapping algorithms on \textit{MSP-Podcast} (with 5 classes) have different performances. 
It reveals the limitation of the proposed idea of using regression models for the classification task.

It should be mentioned that the objective of this study is not to outperform the classification task, but only to compare the performance of the two approaches.
The main idea of this paper is to study the performance of a regression model on AVD space, which can be used for classification as well.
The regression model can use all available information in a corpus (samples are not limited to certain categories) for training.
Moreover, the class label can be an interpretation of the model's output, which means it is not necessary to have a categorical annotation of samples.
This interpretation or mapping from dimensional space to categorical labels can be simply done by defining the emotional classes in continuous space in the posterior.

\section{EXPERIMENT DESIGN}

We propose to build a regression model to predict a vector of values in the continuous space as the representation of the emotional state.
The output of a trained regressor can be fed to a mapping model to transform into emotional labels.
The training of mapping models defines the categorical emotions in dimensional space.
In our experiment, they are constructed using the training set of the corresponding corpus.
% Besides the benefits of analyzing the emotions in a dimensional space, studying the performance of classification via regression model is the main aim of this paper.
The performance of classification via the regressor model (the right pipeline in the Figure \ref{fig_cla_via_reg}) is compared to a classification model (the left pipeline in the Figure \ref{fig_cla_via_reg}).
% The objective of this study is to investigate the potential of the regression model in emotion recognition.
The main aim is not to outperform the state-of-the-art system, but only to propose a different view toward emotion recognition and its potential.
Although, the state-of-the-art methods such as data augmentation \cite{pappagari2020xvec}, reject option \cite{sridhar2019speech}, use fine-tuning or more fancy feature extractors \cite{pepino2021emotion,wang2021fine} can be applied to improve the performances.

Using a similar architecture for the classifier and regressor provides the chance of comparing two approaches with the almost same capacity of learning (number of network's weights).
The regression models employ a linear layer as output and MSE as their loss function.
The classification models are similarly designed, with some modifications. 
Their output layer is adapted to the number of classes, the output layer is modified to the softmax, and cross-entropy is used as their loss function.

In order to create the classifiers and regressors, a combination of three feature extractors and three different neural network architectures has been tested.

\subsection{Feature extractors}

By emerging of pretrained neural network models and their decent performance on different tasks, in particular for emotion recognition \cite{pappagari2020xvec,pepino2021emotion}, we propose to use pretrained \textit{wav2vec2} \cite{baevski2020wav2vec} and \textit{wavLM} \cite{chen2021wavlm} models as the feature extractor.

The wav2vec2 \cite{baevski2020wav2vec} used self-supervised learning on raw audio to transform it into an embedding representation.
We used the \textit{wav2vec 2.0 base} model, pre-trained on Librispeech (960 hours of speech) without fine-tuning.
Same as wav2vec2, the wavLM Base+ \cite{chen2021wavlm} extracts universal
speech representations.
But unlike wav2vec2, the \textit{wavLM} model has been trained on massive unlabeled speech data (94k hours of speech).
By using these two pre-trained models, the raw audios are encoded into a sequence of embeddings with a window length of 25ms and a stride of 20ms.

Moreover, the Mel spectrogram (\textit{MelSpg}), which showed a decent performance in \cite{satt2017efficient}, has been used.
In order to have the same length of feature sequence, the same configuration of the sliding window has been set for \textit{MelSpg}.

These three feature extractors generate a vector for each frame of given audio.
To treat all audio signals with variable lengths, in the same way, a padding/truncation method has been applied to have a fixed length (500 frames from the first 6.9 sec) of features. 
128 features per frame have been extracted using \textit{MelSpg}, while using \textit{wav2vec2} and \textit{wavLM} would return 512 features per frame.

\subsection{Classifier/Regressor architectures}

\begin{figure*}[]
\centering
\includegraphics[width=\textwidth]{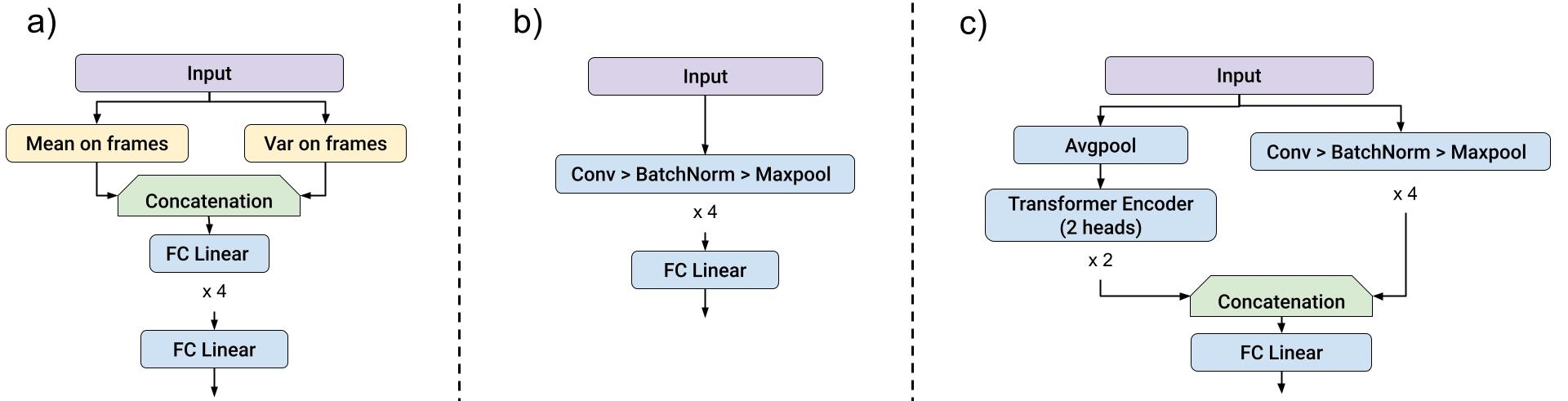}
\caption{Downstream models for regression and classification (by adding a softmax layer as the last layer and adapting the output to the number of classes); (a) MLP model uses the concatenation of mean and variance of features on frames. (b) CNN model is a stacked convolutional layer and max pooling which uses the entire sequence of frames. (c) Parallel CNN - Transformer, which uses the entire sequence of frames.}
% Number of parameters wav2vec2 and wavLM
% MLP : 697,540
% DCNN_M5: 1,333,060
% CNN_Trans: 3,534,494
% Number of parameters MelSpg
% MLP : 304,324
% DCNN_M5: 350,020
% CNN_Trans: 507,806
\label{fig_arch}
\end{figure*}

In order to predict the categorical emotions or the continuous values in AVD space, three different architecture has been designed (see Figure \ref{fig_arch}).

These downstream models are constructed with almost the same number of trainable parameters. 
The \textit{MLP} model (see Figures \ref{fig_arch} - a) constructed with 5 stacked fully connected layers.
In order to generate a prediction, the last layer is a fully connected layer adapted corresponding to the number of classes in the classification task or the number of dimensions (three in AVD) in the regression task.
While the \textit{MLP} model is limited to the mean and variance of frames' features, two other models can profit from the temporal information.
The \textit{CNN} model, inspired by \cite{pepino2021emotion}, (see Figures \ref{fig_arch} - b) is consecutive feed forward of 5 blocks of convolutional layer, batch normalization layer and max polling. 
The \textit{CNN-Trans} model (see Figures \ref{fig_arch} - c) is a parallel downstream architecture.
% https://github.com/Data-Science-kosta/Speech-Emotion-Classification-with-PyTorch
On one side, it is 4 stacked convolutional layers. 
On the other side, the sequential information would be passed through an average pooling to reduce the dimensionality, afterward, it would be fed to two transformer encoder blocks with two heads of attention.
% Then calculated embedding would be passed through a dense layer.
Then the Transformer embedding will be concatenated to the CNN side.
Finally, the result will pass through a fully connected layer to generate the prediction in the same way as the last layer of \textit{MLP}.  

\subsection{Different datasets different challenges}
\label{sec_datasets}

Besides the challenges of different definitions and the consistency of two representation approach  (mentioned in the section \ref{sec_challenge}), the context of speech emotion recognition application plays a major role in the evaluation metrics.
The context of application makes the designing of speech emotion corpora different.
While \textit{IEMOCAP} dataset \cite{busso2008iemocap} is recording of acted (some may call exaggerated) emotions with respecting the frequency of each class, the \textit{MSP-Podcast} dataset \cite{lotfian2019msp} is extracted from recorded spontaneous speech without considering balancing the classes. 

In this study, we use \textit{IEMOCAP} and \textit{MSP-Podcast} datasets to investigate our hypothesis on different contexts.
The \textit{IEMOCAP} dataset with 12h is smaller than the \textit{MSP-Podcast} datasets with 27h of speech.
In order to follow the previous studies \cite{pappagari2020xvec,pepino2021emotion},
% \cite{tarantino2019self,pappagari2020xvec,pepino2021emotion}, 
four main emotions (Natural 30.9\%, Happy 29.6\%, Angry 19.9\%, Sad 19.6\%) in \textit{IEMOCAP} has been selected which makes 5531 samples from all sessions.
Contrary to this balanced distribution of emotions, for the MSP-Podcast datasets, like \cite{lotfian2019curriculum,pappagari2020xvec}, five main emotion classes (Natural 53.3\%, Happy 29.3\%, Angry 6.6\%, Sad 5.4\%, Disgust 5.3\%) have been selected which contains 48754 samples.
The impact of imbalanced classes in the \textit{MSP-Podcast} can be observed in the performance of dummy classifiers in table \ref{tab_dummy}.

% The main performance metrics for evaluation of classifier performance are accuracy or f1-score.
% In some applications, the frequency of classes as weight should be taken into account for any evaluation, especially in imbalanced data.
% It means when the task is emotion recognition in a real-life situation, the weighted performance can be more important. 
% On the other hand, when the task is only to distinguish between different emotions, the unweighted performance can be applied.
% For this purpose, it is decided to report unweighted recall (UR) and weighted recall (WR).
% Needless to mention, contrary to classification models, the problem of imbalanced data is necessary to be taken care of in regression models.

% 5fold cross-validation for IEMOCAP vs original partitioning of MSP
The reported performance in this study is based on 5-fold cross-validation for the \textit{IEMOCAP} dataset.
The original partitioning of the \textit{MSP-Podcast} dataset into train/dev/test is respected, and evaluations are based on the sum of test1 and test2 partitions \cite{lotfian2019msp}.

% \cite{pappagari2020x} : 58.46\%

% \cite{pappagari2020x} : 70.30\%

% \cite{tarantino2019self}: WA:64.33\% UA:64.79\%

\section{RESULTS}

The performance of regressor models (three feature types, three architectures) has been calculated based on the Concordance Correlation Coefficient (CCC) \cite{weninger2016discriminatively}.
The output of the regressor model as AVD values has been mapped to categorical emotions using three algorithms mentioned in the section \ref{sec_class_via_reg}.
The classification performance of these two approaches with different configurations on the \textit{IEMOCAP} dataset and \textit{MSP-Podcast} test set, are shown in tables \ref{tab_iemocap_res} and \ref{tab_msp_res}.

\begin{table*}[t]
\centering
\resizebox{\textwidth}{!}{%
\begin{tabular}{ll|cc|c|ccc}
\hline
\multicolumn{2}{c|}{Model}     &    \multicolumn{2}{|c|}{Classification} & \multicolumn{1}{|c|}{Regression} & \multicolumn{3}{c}{Classification via Regression (UR, WR)}    \\
\hline
 Feature   & Architecture     &    UR &    WR & CCC (A, V, D)             & 2LP    & KNN    & Gaussian\\%(noP)   & GNB(P) \\
\hline
 MelSpg      & MLP       & 25.02 & 26.00 & (0.15, 0.65, 0.46) & (35.56, 40.03) & (34.49, 38.84) & (38.30, 41.81) \\%& (36.28, 40.55)  \\
 MelSpg      & CNN   & 54.14 & 53.73 & (0.36, 0.66, 0.50)  & (44.86, 48.00)  & (42.81, 46.17) & (45.38, 47.79) \\%   & (43.53, 47.01)  \\
 MelSpg      & CNN-Trans & 56.60 & 56.13 & (0.26, 0.66, 0.50)  & (43.33, 45.21) & (42.07, 44.57) & (45.06, 46.05)\\%    & (43.29, 45.58)  \\
%  wav2vec2  & MLP       & \textbf{59.16} & \textbf{58.88} & \textbf{(0.48, 0.72, 0.56)} & (50.29, 52.57) & (48.70, 51.64)  & \textbf{(52.21, 53.08) }\\%    & (49.82, 52.58)  \\
 wav2vec2  & MLP       & \textbf{63.16} & \textbf{61.81} & \textbf{(0.44, 0.72, 0.57)} &  (49.63, 51.40) &  (47.07, 48.63) & \textbf{(51.44, 53.00) } \\%    & (49.82, 52.58)  \\
 wav2vec2  & CNN   & 31.93 & 35.79 & (0.10, 0.42, 0.32)  & (32.48, 36.37) & (30.83, 34.70)  & (34.56, 37.82)\\%    & (32.92, 36.87)  \\
%  wav2vec2  & CNN-Trans & 58.01 & 57.12 & (0.23, 0.48, 0.39) & (39.92, 42.38) & (38.30, 41.06)  & (40.73, 42.40)\\%     & (38.41, 41.12)  \\
 wav2vec2  & CNN-Trans & 61.42 & 59.16 & (0.31, 0.67, 0.52) & (45.35, 47.57) & (43.06, 45.93) & (46.74, 48.17)\\%     & (38.41, 41.12)  \\
 wavLM     & MLP       & 57.61 & 57.26 & (0.37, 0.72, 0.55) & (47.48, 50.44) & (45.55, 48.85) & (50.05, 52.10)\\%     & (46.99, 50.12)  \\
 wavLM     & CNN   & 35.38 & 38.41 & (0.04, 0.24, 0.20)  & (27.59, 32.27) & (27.99, 32.51) & (28.10, 32.40)\\%      & (27.49, 32.15)  \\
 wavLM     & CNN-Trans & 57.78 & 55.95 & (0.19, 0.47, 0.38) & (40.05, 42.14) & (37.61, 40.18) & (40.76, 42.14)\\%    & (38.25, 40.78)  \\
\hline
\end{tabular}
}
\caption{Classification and Regression result on the \textit{IEMOCAP} (four emotional classes)}
  \label{tab_iemocap_res}
\end{table*}

\begin{table*}[t]
\centering
\resizebox{\textwidth}{!}{%
\begin{tabular}{ll|cc|c|ccc}
\hline
\multicolumn{2}{c|}{Model}     &    \multicolumn{2}{|c|}{Classification} & \multicolumn{1}{|c|}{Regression} & \multicolumn{3}{c}{Classification via Regression (UR, WR)}    \\
\hline
 Feature   & Architecture     &    UR &    WR & CCC (A, V, D)             & 2LP    & KNN    & Gaussian\\%(noP)   & GNB(P) \\
\hline
 MelSpg      & MLP       & 20.00 &  50.60 & (0.30, 0.05, 0.21)  & (20.04, 51.35) & (20.16, 51.48) & (20.47, 51.72)\\%    & (20.21, 51.55)  \\
 MelSpg      & CNN   & 28.34 & 54.88 & (0.41, 0.09, 0.31) & (22.87, 53.62) & (22.92, 53.39) & (23.45, 54.04)\\%    & (22.50, 53.89)   \\
 MelSpg      & CNN-Trans & 29.08 & 54.10 & (0.43, 0.10, 0.30)   & (22.80, 54.60)   & (22.54, 54.34) & (23.85, 55.14)\\%    & (24.56, 55.15)  \\
 wav2vec2  & MLP       & \textbf{36.89} & 57.92 & \textbf{(0.51, 0.17, 0.40)}  & (21.87, 53.62) & (21.34, 53.05) & \textbf{(25.88, 56.25)}\\%    & (23.90, 55.68)   \\
 wav2vec2  & CNN   & 26.42 & 51.36 & (0.13, 0.01, 0.09) & (20.02, 51.30)  & (20.09, 51.38) & (20.39, 51.55)\\%    & (20.10, 51.36)   \\
 wav2vec2  & CNN-Trans & 33.98 & \textbf{58.76} & (0.46, 0.06, 0.37) & (20.41, 51.86) & (20.21, 51.54) & (23.46, 55.24)\\%    & (21.67, 53.47)  \\
 wavLM     & MLP       & 34.87 & 57.14 & (0.49, 0.13, 0.39) & (21.23, 53.12) & (20.79, 52.50)  & (23.47, 55.89)\\%    & (23.31, 55.73)  \\
 wavLM     & CNN   & 27.39 & 52.70 & (0.00, -0.00, 0.01)  & (20.00, 51.28)  & (20.00, 51.28)  & (20.01, 51.24)\\%    & (20.01, 51.30)   \\
 wavLM     & CNN-Trans & 33.05 & 56.89 & (0.41, 0.03, 0.32) & (20.15, 51.52) & (20.21, 51.61) & (23.28, 55.52)\\%    & (23.31, 55.41)  \\
\hline
\end{tabular}
}
\caption{Classification and Regression result on the \textit{MSP-Podcast} (five emotional classes)}
  \label{tab_msp_res}
\end{table*}

The results on two corpora show a low performance of classification if temporal information would not be taken into account. 
Using \textit{MelSpg} in \textit{MLP} architecture achieves a lower performance (almost equal to a dummy performance in \textit{MSP-Podcast}) comparing with cases which temporal information have not been used in extracted features (\textit{wav2vec2} or \textit{wavLM}) or in the model architectures (\textit{CNN} or \textit{CNN-Trans}).

In the \textit{IEMOCAP} dataset (see table \ref{tab_iemocap_res}), the best classification results obtained by feeding the extracted embedding from \textit{wav2vec2} into \textit{MLP} model with UR=63.16\% and WR=61.81\%.
The same configuration achieved the best regression performance with the CCC of Arousal equal to 0.44, the CCC of Valence equal to 0.72, and the CCC of Dominance equal to 0.57.
The best performance of classification via regression is obtained by applying \textit{Gaussian} classifier as the mapping algorithm.
Our experiment shows a lower performance of Gaussian Naive Bayes algorithm, which uses the frequency of classes in the training set as the prior probability.
Comparing "classification" and "classification via regression" approaches shows that using dimensional annotation can successfully classify emotions with more than 80\% proportional performance to using direct categorical annotation (UR=51.44\% and WR=53.00\%).

The best performances achieved using \textit{wav2vec2} or \textit{wavLM} features and \textit{MLP} or \textit{CNN-Trans} models, which is decent compare to \cite{pappagari2020xvec}.
Same as \textit{IEMOCAP} corpus, the best performance of regression and classification via regression obtained using \textit{wav2vec2} and \textit{MLP}.
We believe that there are several reasons for poor performance of models on the \textit{MSP-Podcast} dataset, table \ref{tab_msp_res}. 
As it has been observed in the table \ref{tab:kri_a}, the annotator agreement is low in this corpus. 
Also, the emotions are not acted, which makes them more difficult to recognize.
Moreover, most of the samples in this corpus are labeled as Natural, which makes classification more difficult in imbalanced problems (see poor UR in table \ref{tab_msp_res}).

The performance of classification via regression on these two corpora shows the potential and the limitation of this approach.
Although following this approach can reduce the performance of classification, there are still several advantages.
Using only dimensional annotation for training a regressor model can reduce the cost of preparing a dataset and using all available data.
It also can prepare a model which can be used for the recognition of new emotion categories by only providing its definition in the AVD space. 
In order to show the potential of "classification via regression", we trained an \textit{MLP} model using \textit{wav2vec2} features on samples from \textit{IEMOCAP} with 4 emotions (Natural, Happy, Angry, Sad) as a regressor.
Then we asked the pipeline to classify emotions with a new test set containing Frustration emotion, which change the problem from four to five classes.
The model is able to recognize the Frustration class with a precision of 41.85\% and recall of 33.38\%, although it has not seen any speech sample of this class in the regressor's training.

% Comparing with state of the art systems:
% weighted f-scores 
% reported in \cite{pappagari2020xvec} : 
% IEMOCAP: 65.95\% MSP-Podcast: 57.42\%
% Our : IEMOCAP: 61.21\% MSP-Podcast: 51.58\%

% WA on Balanced test in "SUPERB: Speech processing Universal PERformance Benchmark":
% reported in \cite{pappagari2020xvec} :  IEMOCAP: 67.62\%
% Our IEMOCAP: 63.98\%

\section{CONCLUSION}

%problematics
In this work, we investigated the relationship between dimensional representation and categorical labeling of emotions.
We proposed to consider the speech emotion recognition task as a regression problem whose output can be interpreted as categorical emotions by using a mapping algorithm. 

% experiments and results
However, there are several benefits of following the "classification via regression" approach, our experiment on two different corpora shows degradation of performance compared to a classifier that profits from categorical labeled data.
We compared the performance of these two approaches by employing three feature types, three architectures as regressor/classifier, and three mapping algorithms for transforming the continuous value in the AVD space to categorical labels.
While the main objective of this study was not to outperform the state of the art, we showed the potential and limitations of the "classification via regression" approach.

% perspective
The benefits of the proposed approach are reducing the cost of data preparation and breaking the limitation of the predefined number of emotional categories.
The performance of the model can be improved by applying the state-of-the-art techniques, such as fine-tuning the pre-trained \textit{wav2vec2} or \textit{wavLM} feature extractors \cite{pepino2021emotion,wang2021fine}.
Although the performance of the best regressor model and mapping algorithm as the classifier is decent, particularly in the \textit{IEMOCAP} dataset.

The importance of dimensions in predicting categorical labels (see Section \ref{sec_class_via_reg}) suggests modifying the loss function of the regressor model to a weighted loss in the future works.

%%%%%%%%%%%%%%%%%%%%%%%%%%%%%%%%%%%%
%%%%%%%%%%%% TODO %%%%%%%%%%%%%%%%%%
% data augmentation with wav2vec2 and MLP
% Defining wiegthed CCC as loss function for regression model \cite{atmaja2020dimensional,li2021contrastive} 
%%%%%%%%%%%%%%%%%%%%%%%%%%%%%%%%%%%%

% \section{REFERENCES}
% \label{sec:ref}

% List and number all bibliographical references at the end of the
% paper. The references can be numbered in alphabetic order or in
% order of appearance in the document. When referring to them in
% the text, type the corresponding reference number in square
% brackets as shown at the end of this sentence \cite{C2}. An
% additional final page (the fifth page, in most cases) is
% allowed, but must contain only references to the prior
% literature.

% References should be produced using the bibtex program from suitable
% BiBTeX files (here: strings, refs, manuals). The IEEEbib.bst bibliography
% style file from IEEE produces unsorted bibliography list.
% -------------------------------------------------------------------------
% \bibliographystyle{IEEEbib}
% \bibliography{strings,refs}
\bibliographystyle{unsrt}  
\bibliography{Template_Blind}  

\end{document}